\documentclass[]{aastex}
\usepackage{emulateapj5}
\usepackage{onecolfloat}
\usepackage{graphicx} 
\usepackage{fancyheadings} 
\usepackage{ulem}
\usepackage{rotating}
\usepackage{lscape}

\newcommand{\napj}{Astrophys.\ J.}
\newcommand{\napjl}{Astrophys.\ J. Lett.}
\newcommand{\napjs}{Astrophys.\ J. Supp.}
\newcommand{\napjsupp}{Astrophys.\ J. Supp.}
\newcommand{\naap}{Astron.\ Astrophys.}
\newcommand{\naj}{Astron.\ J.}
\newcommand{\npasp}{Proc.\ Astron.\ Soc.\ Pacific}
\newcommand{\naraa}{Ann.\ Rev.\ Astron.\ Astrophys.}

\newcommand{\lya}{Ly$\alpha$ }

\newcommand{\cm}[1]{\, {\rm cm^{#1}}}
\newcommand{\N}[1]{{N({\rm #1})}}

\newcommand{\mkms}{{\rm \; km\;s^{-1}}}

\begin{document}

\twocolumn[%
\submitted{To appear in Nature: April 2003}

\title{The elemental abundance pattern in a galaxy at $z=2.626$}

\author{ JASON X. PROCHASKA}
\affil{UCO/Lick Observatory}
\affil{University of California, Santa Cruz;
Santa Cruz, CA 95064 (xavier@ucolick.org)}

\and

\author{J. CHRISTOPHER HOWK \& ARTHUR M. WOLFE}
\affil{Department of Physics, and Center for Astrophysics and Space Sciences}
\affil{University of California, San Diego; 
C--0424; La Jolla, CA 92093}

{\bf 
The discovery of metal-poor stars$^{1,2}$ (where metal is any element
more massive than helium) has enabled astronomers to probe the chemical
enrichment history of the Milky Way$^{3,4}$.
More recently, element abundances in gas inside high-redshift galaxies
has been probed through the absorption lines imprinted on the spectra
of background quasars$^{5-8}$, but these have typically yielded
measurements of only a few elements.  Furthermore, interpretation
of these abundances is complicated by the fact that differential
incorporation of metals into dust can produce an abundance pattern  
similar to that expected from nucleosynthesis by massive stars$^9$.
Here we report the observation of over 25 elements in a galaxy at
$z=2.626$. 
With these data,  we can examine nucleosynthetic 
processes independent of the uncertainty arising from depletion.
We find that the galaxy was enriched mainly by
by massive stars ($M > 15$ solar masses)
and propose that it is the progenitor of a massive, elliptical galaxy.
The detailed abundance patterns suggest that boron is produced
through processes that act independently of metallicity,
and may require alternative mechanisms for the nucleosynthesis of   
germanium.
}
]

\pagestyle{fancyplain}
\lhead[\fancyplain{}{\thepage}]{\fancyplain{}{PROCHASKA, HOWK \& WOLFE}}
\rhead[\fancyplain{}{The elemental abundance patter in a galaxy at $z=2.626$}
]{\fancyplain{}{\thepage}}
\setlength{\headrulewidth=0pt}
\cfoot{}

With the exception of the light elements (Li, Be, B, and partly He) whose
origin is linked to either big bang nucleosynthesis$^{10}$ or unique
astrophysical processes$^{11}$, the elements through Fe are produced 
during the nuclear reactions which fuel stars or during the collapse
of the stellar core and the resulting explosion defining a
supernovae (SN)$^{12}$.  
Regarding supernovae, the current physical picture
states that even-Z nuclei including the '$\alpha$-elements'
O, Ne, Mg, Si, S, Ar, and Ca are synthesized by the SN of massive stars
whereas a substantial component of the
Fe-peak elements are produced on significantly longer time-scales.
in SN with lower mass progenitors.
Therefore, comparisons between the $\alpha$-elements and the Fe-peak
reflect on the star formation history and age of the galaxy$^{13}$.
The heavier elements -- whose nuclei are energetically less 
favorable than Fe --
require a significant source of free neutrons and the sequence of
elements and isotopes created is very sensitive to the neutron capture 
rate relative to $\beta$-decay: the s-process (r-process) refers to a slow 
(fast) rate.  To date, these heavier elements have never been detected
outside our Galaxy and its nearest neighbors.

The galaxy central to this Letter was discovered$^{14}$ 
in absorption at redshift $z=2.626$ along the sightline to the 
background quasar
FJ081240.6+320808 (emission redshift $z_{em} = 2.701$) 
via the signature damping wings
of the \lya profile which give the galaxy its classification 
-- a damped \lya system$^{15}$ (DLA).
We obtained a moderate resolution spectrum of FJ081240.6+320808
with the Echellette Spectrograph and Imager$^{16}$.
on the Keck~II telescope as part of a larger program to survey the
chemical enrichment history of the universe$^{17}$.
The unusually strong
metal-line absorption profiles of this galaxy 
led us to acquire follow-up observations of FJ081240.6+320808
with the HIRES echelle spectrograph$^{18}$ on Keck~I.  
Our observations revealed a series of metal-line transitions (Figure~1)
previously undetected at high redshift which provide the most comprehensive
set of elemental abundances to date.
By performing standard line-profile fitting and equivalent
width measurements, we have measured gas-phase abundances for the 
elements listed in Table~1.  This DLA has the highest combined H\,I
column density and metallicity (O/H $\approx 1/3$ solar abundance) 
ever observed.
Remarkably, its redshift implies a strict upper limit to its
age of 2.5~Gyr in any realistic cosmology.  

The gas-phase abundances listed in column~2 of Table~1
qualitatively follow the differential depletion pattern observed
along the sightlines through the interstellar medium of the
Milky Way galaxy$^{19}$.   Refractory elements like
Fe, Ni, and Cr must be highly depleted onto dust grains in this high
redshift galaxy.
The presence of strong C\,II$^*$ and Cl\,I absorption suggests the
gas resides in a cold neutral medium characteristic of highly depleted
gas in the Milky Way.  Furthermore,
the observation of significant Cl\,I 
requires at least a modest molecular hydrogen fraction$^{20}$.
To examine the underlying enrichment history of this galaxy we have applied
empirical, {\it conservative} dust corrections (column 3) derived
from the depletion patterns of the Milky Way
to the gas-phase abundances.
Although these corrections are based on our limited
knowledge of dust in the local universe, we stress that the following
discussion of nucleosynthesis is not sensitive to the corrections
unless otherwise noted.

\begin{table}[h]\footnotesize
\begin{center}
\caption{{\sc CHEMICAL ABUNDANCES\label{tab:dust}}}
\begin{tabular}{lccccccccccc}
\tableline
\tableline
Elm & [X/H]$^a$ & $\delta_{dust}^b$ & [X/O]$^c$ \\
\tableline
B &$-0.57$&$ 0.1$&$-0.03$ \\  
N &$>-2.24$&$ 0.0$&$>-1.80$ \\  
O &$-0.54$&$ 0.1$&$ 0.00$ \\  
Mg&$-0.78$&$ 0.3$&$-0.04$ \\  
Al&$>-2.00$&$> 0.5$&$>-1.06$ \\  
Si&$-0.91$&$ 0.3$&$-0.17$ \\  
P &$<-1.06$&$< 0.3$&$<-0.32$ \\  
S &$-0.87$&$ 0.1$&$-0.33$ \\  
Cl&$-1.55$&$> 0.0$&$>-1.11$ \\  
Ti&$-1.87$&$> 0.7$&$>-0.73$ \\  
Cr&$-1.61$&$> 0.7$&$>-0.47$ \\  
Mn&$<-1.85$&$ 0.7$&$<-0.71$ \\  
Fe&$-1.69$&$> 0.7$&$>-0.55$ \\  
Co&$<-1.48$&$> 0.7$&$>-0.34$ \\  
Ni&$-1.73$&$> 0.7$&$>-0.59$ \\  
Cu&$<-1.11$&$> 0.7$&$> 0.03$ \\  
Zn&$-0.91$&$ 0.2$&$-0.27$ \\  
Ga&$<-1.45$&$ 0.7$&$<-0.31$ \\  
Ge&$-0.92$&$ 0.3$&$-0.18$ \\  
As&$< 0.26$&$ 0.0$&$< 0.70$ \\  
Kr&$<-0.44$&$ 0.0$&$< 0.00$ \\  
Sn&$<-0.27$&$ 0.0$&$< 0.17$ \\  
Pb&$<-0.10$&$ 0.0$&$< 0.34$ \\  
\tableline
\end{tabular}
\tablenotetext{a}{Gas-phase abundances relative to solar on a logarithmic scale, 
e.g., [X/H] = --1 implies 1/10 solar abundance.
Throughout the analysis we have adopted $\N{HI} = 10^{21.35} \cm{-2}$
measured from a fit to the damped \lya profile.}
\tablenotetext{b}{Dust corrections estimated from the depletion patterns of Galactic 
gas with comparable depletion levels$^{19}$. The values are added to
the gas-phase abundances to yield the underlying nucleosynthetic pattern. 
In all cases, we have considered very conservative corrections.}
\tablenotetext{c}{Dust-corrected abundances on a 
solar logarithmic scale relative to O. 
These values express the nucleosynthetic pattern of this protogalaxy.}
\end{center}
\end{table}

Figure~2 reveals the nucleosynthetic pattern of this galaxy 
where the dotted-line compares the solar abundance pattern scaled to the
oxygen metallicity of the galaxy.  At the crudest level, the galaxy's
enrichment pattern resembles the Sun; this point is both comforting
and impressive given its age and distinction from our current universe.
Quantitatively, one might expect that the galaxy's young age and 
relatively high metallicity would imply a nucleosynthetic pattern
dominated by short-lived, massive stars.
This presumption is supported by several lines of evidence.
First, the $\alpha$-elements (O,Mg,Si)
exhibit enhanced relative abundances compared to Zn, 
the only element representative of the Fe-peak not severely incorporated
into dust grains. 
This $\alpha$/Zn enhancement is suggestive of enrichment by massive stars.
Second, the relative abundances of the $\alpha$-elements exhibit a trend
of lower relative abundance at higher atomic number Z 
(e.g.\ [O/S]~$\approx 0.3$).  This trend
reflects the detailed production factors predicted for 
very massive stars$^{21}$.
Finally, all of the odd-Z elements show sub-solar relative
abundances marking the enhanced 'odd-even effect' indicative
of massive SN.  This includes
[Mn/Fe]~$< -0.16$, [Ga/Ge]~$< -0.13$, and [P/Si]~$< -0.15$.

The high metallicity and relative abundances of this galactic gas
resemble stellar abundances observed in massive, early-type galaxies 
(e.g.\ ellipticals) dominated by old stellar populations$^{22}$. 
One might speculate, then, that this galaxy is the progenitor of one
of these massive structures which could explain its high
metallicity at such an early epoch.
In fact, 
its metallicity and dust content resemble those for 
the Lyman break galaxies$^{23}$, galaxies identified in emission at $z>3$
which are the presumed progenitors of massive galaxies.
The similarity in redshift, metallicity, and relative abundances
with the well-studied Lyman break galaxy cB58
is particularly note-worthy$^{24}$.
Furthermore, early-type galaxies may exhibit$^{25}$
an overabundance
of Mg/Ca which resembles the $\alpha$-element vs.\ Z trend 
observed for this protogalaxy.  
Follow-up imaging and spectroscopy
of the galaxies in the field surrounding FJ081240.6+320808
may help reveal its stellar properties.

Our data place unique constraints on nucleosynthetic processes in the
early universe.  
The observations, for example, help distinguish between
processes where B production scales with the galaxy's metallicity and 
processes that are independent of metallicity.  The former theories
(e.g.\ neutrino spallation in the carbon shells of SN$^{26}$;
the spallation of C and O nuclei accelerated by SN onto 
local interstellar gas$^{27}$) predict the B/O ratio
remains constant with O/H metallicity while the latter 
(e.g.\ p,n accelerated onto interstellar CNO seed nuclei) 
predict the B/O ratio increases with increasing metallicity.
The current observation of a solar B/O ratio in this $\approx 1/3$ solar
metallicity gas argue for B synthesis independent of metallicity.
Future observations will test this hypothesis in other galaxies
and may allow a measurement of the $^{10}$B/$^{11}$B  
isotopic ratio critical to determining the 
relative importance of neutrino and cosmic ray spallation$^{12}$. 

The impact of our observations is particularly powerful when
interpreted under the light of the young age of this galaxy.  
For example, the nearly solar Ge/O ratio raises new 
challenges for the production of Ge. 
Although the s-process is expected to be a 
principal channel for Ge nucleosynthesis, the primary s-process site 
is during the 'asymptotic giant branch' phase of
low mass stars$^{28}$ whose lifetimes may exceed the age of this galaxy.
In addition to the s-process, SN with low mass progenitors may produce
a significant fraction of the Ge observed in our Galaxy.
We have argued, however, that this high redshift galaxy is 
dominated by enrichment
from massive stars, therefore the observed Ge/O ratio may argue
for a distinct physical processes (e.g.\ the neutrino-wind process$^{29}$).
These and related issues will be investigated 
by detecting or significantly lowering the current limits 
for Pb, Sn, and Ga; Pb and Sn production are dominated by the s-process
while Ga may be produced in a similar fashion to Ge.

This galaxy is also special for 
exhibiting transitions of C\,I, C\,I$^*$, C\,I$^{**}$, Cl\,I,
C\,II$^*$, and Si\,II$^*$ which will yield an unparalleled analysis
of its physical properties$^{30}$ (e.g.\ $\rho$, $T$, pressure, star
formation rate) and
an assessment of the cosmic microwave background temperature at $z=2.6$.
These observations will enable us to characterize
the physical characteristics and star formation history of this young
galaxy as well as provide detections or important limits on
Ar, Ti, C, N, P, Pb, Sn, and possibly Kr and F.  
In turn, we will test theories related to CNO and r-process 
nucleosynthesis.

Absorption-line studies of DLA provide the only means of examining 
elements beyond the Fe-peak outside the very local universe.
The discovery of this galaxy indicates similar abundance analyses
will be possible in a small but non-negligible sample 
($\approx 2\%$) of high redshift DLA.
In closing, we report the discovery of a second DLA which should present
an abundance analysis of $\approx 20$ elements
including Ga, Ge, Cu and possibly B.
Remarkably, this DLA is identified along the same 
sightline to FJ081240.6+320808 but at a cosmologically distinct redshift.

\acknowledgments

{\small These observations were made with the W.M. Keck Telescope.
The Keck Observatory is a joint facility of the University
of California, the California Institute of Technology, and NASA.
The author wishes to recognize and acknowledge the very significant cultural
role and reverence that the summit of Mauna Kea has always had within the
indigenous Hawaiian community.  We are most fortunate to have the
opportunity to conduct observations from this mountain.
JCH recognizes support from a NASA grant to UC San Diego.

}

\begin{figure*}
\begin{center}
\includegraphics[width=6.5in]{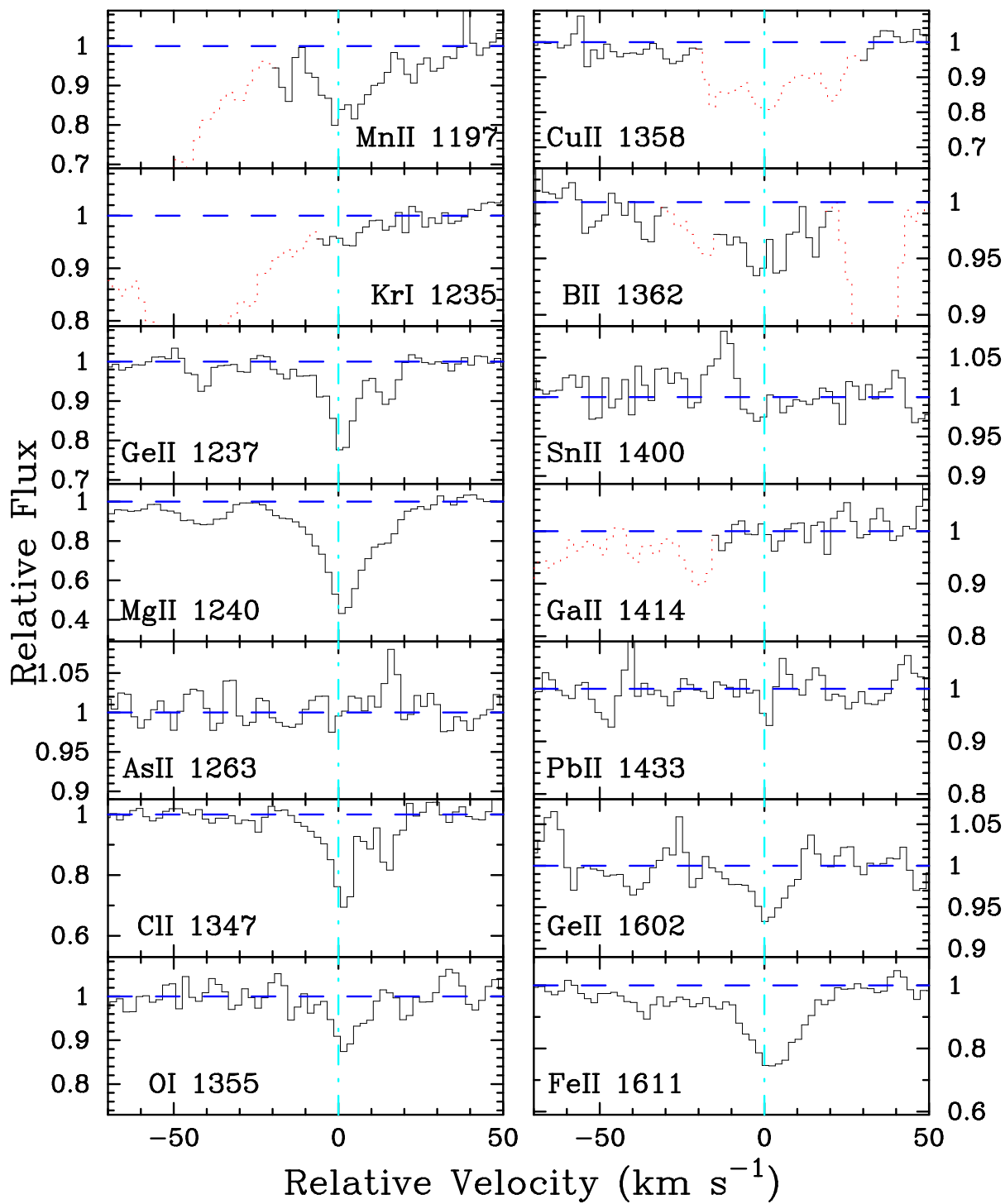}
\caption{{\bf Sample of previously undetected metal-line transitions:}
Normalized absorption-line profiles related to the galaxy
at $z=2.626$ toward FJ081240.6+320808 taken with the HIRES echelle
spectrograph on the Keck~I 10m telescope ($R \approx 45000$, 
S/N $\approx 30$ per 2~km/s pixel).  
With the exception of Fe\,II 1611 and
Mg\,II 1240, none of these transitions have been
detected outside of our Galaxy.  Line blends with coincident absorption
features are designated by dotted lines.  The velocity $v = 0 \mkms$ 
was arbitrarily defined to correspond to $z=2.6263$.   
With the exception of Cl\,I, these transitions correspond to the
dominant ion of these elements in neutral hydrogen gas. 
Therefore, 
we convert the ionic column densities measured from these transitions
into elemental abundances without ionization corrections. 
Although several transitions provide only upper limits
to the elemental abundance (e.g.\ Pb\,II 1433, Ga\,II 1414), the
spectral regions are free from line-blends and future observations
will yield detections or important limits.
}
\label{fig:mtl}
\end{center}
\end{figure*}

\begin{figure*}[ht]
\begin{center}
\includegraphics[width=5.5in,angle=-90]{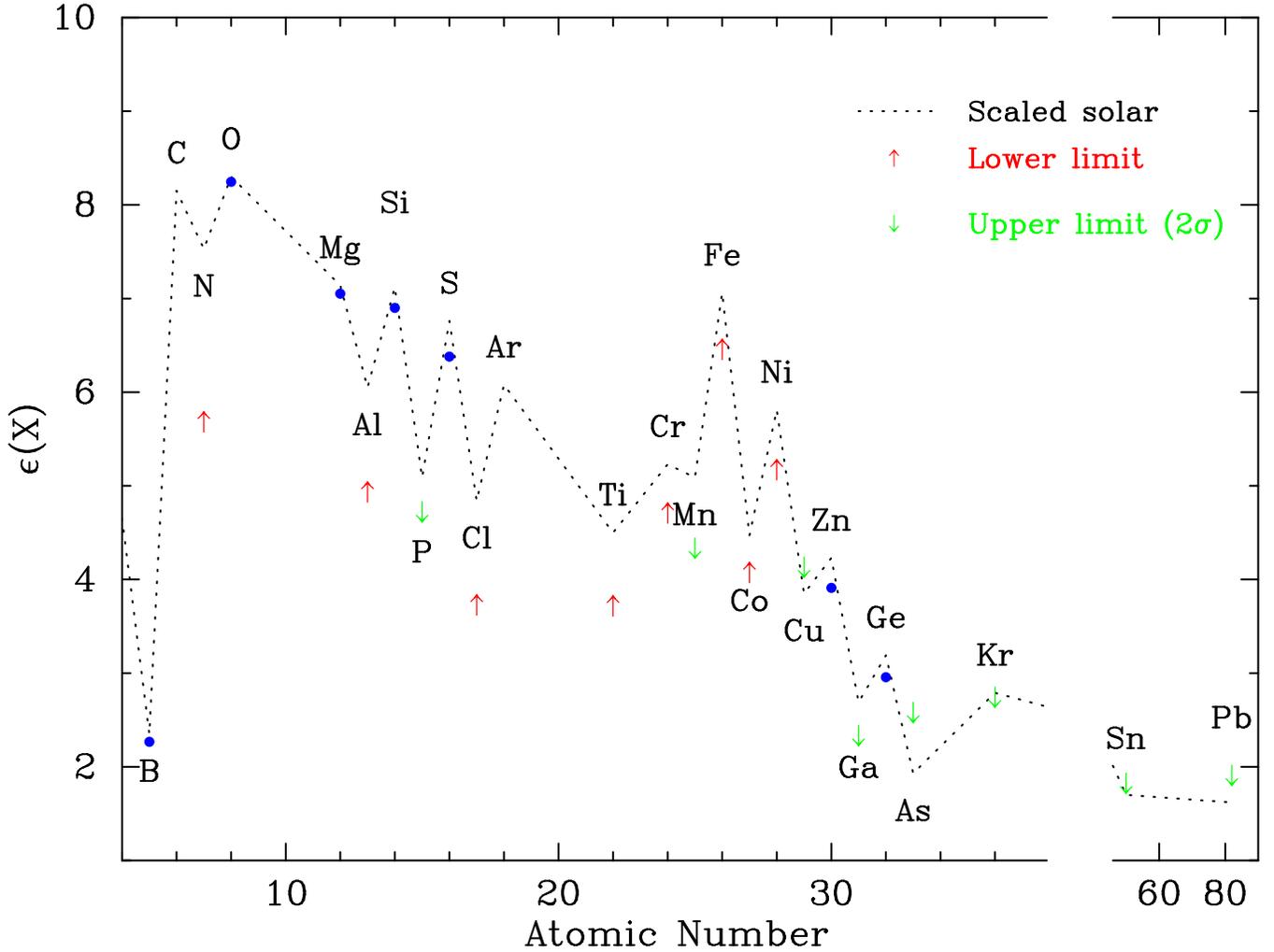}
\caption{{\bf The nucleosynthetic enrichment pattern for a galaxy discovered
in the early universe:}
Dust corrected elemental abundances for the protogalaxy at $z=2.626$
toward FJ081240.6+320808 on a logarithmic scale where hydrogen 
is defined to have $\epsilon (X) = 12$.  
The dust corrections were empirically derived from depletion patterns observed 
in the local universe and were applied in a very conservative manner.  
This explains the lower limits to the Fe, Cr, Al, Co, and Ni 
abundances and similarly the upper limit to Mn.   
Typical statistical errors are $< 0.1$~dex, i.e., the size of the 
plot symbols. 
The dotted line traces the solar abundance pattern
scaled to match the observed Oxygen abundance ([O/H] = --0.44, after
dust correction).
To zeroth order, the pattern of this high redshift galaxy resembles
that of our Sun indicating their nucleosynthetic enrichment histories
are not too dissimilar.  At finer detail, one notes several important
differences:  (i) (O,Mg,Si) are enhanced relative to Zn and therefore
presumably to the entire Fe-peak.  This enhancements is indicative of
enrichment by massive stars; (ii) The $\alpha$-elements show lower
relative abundance at higher Z (e.g.\ [O/S]~$\approx -0.3$), matching
the detailed production factors of nucleosynthesis in massive stars and their SN;
(iii) there is an enhanced 'odd-even effect' with P, Ga and Mn showing
sub-solar relative abundances relative to Si, Ge, and Fe. 
Finally, we emphasize the solar B/O and nearly solar Ge/O ratios.
The former observation argues against the 'secondary' production
of B which predict the B/O ratio scales with metallicity and for
'primary' B production. 
Meanwhile, the Ge/O ratio may challenge current theories for the
production of Ge particularly when one considers the young age
(less than 2.5~Gyr) of the galaxy and the likelihood that it
was primarily enriched by massive stars.
}
\label{fig:all}
\end{center}
\end{figure*}

\clearpage

{\bf References}

{\tiny

\begin{enumerate}

\item Chamberlain, J.W., \& Aller, L.H. 
The atmospheres of A-type subdwarfs and 95 Leonis.
\napj, 114, 52-72 (1951)

\item Helfer, H.L., Wallerstein, G., \& Greenstein J.L.
Abundances in some Population II K Giants.
\napj, 129, 700-719 (1959)

\item Wheeler, J.C., Sneden, C., \& Truran, J.W.Jr.
Abundance ratios as a function of metallicity.
Ann.\ Rev.\ Astron.\ Astrophys., 27, 279-349 (1989)

\item McWilliam, A., Preston, G.W., Sneden, C., \& Searle, L. 
A spectroscopic analysis of 33 of the most metal-poor stars.I.
\naj, 109, 2757-2799 (1995)

\item Lu, L., Sargent, W.L.W., Barlow, T.A.,
Churchill, C.W., \& Vogt, S. 
Abundances at high redshifts: The chemical enrichment 
history of damped Ly$\alpha$ galaxies. \napjsupp, 107, 475-519 (1996)

\item Pettini, M., Ellison, S.L., Steidel, C.C., Shapely, A.L., \& Bowden, D.V.
Si and Mn Abundances in damped Ly$\alpha$; systems with low dust content.
\napj, 532, 65-76 (2000)

\item Molaro, P., Bonifacio, P., Centuri${\rm \acute o}$n, M.,
D'Odorico, S., Vladilo, G., Santin, P., \& Di Marcantonio, P.
UVES observations of QSO 0000-2620: Oxygen and Zinc abundances in the damped
Ly$\alpha$ galaxy at zabs=3.3901. \napj, 541, 54-60 (2000)

\item Prochaska, J.X. \& Wolfe, A.M.,
The UCSD HIRES/Keck I damped Ly$\alpha$ abundance Database. II. 
The implications. \napj, 566, 68-92 (2002)

\item Vladilo, G. Chemical abundances of damped Ly$\alpha$ systems:
A new method for estimating dust depletion effects.
\naap, 391, 407-415 (2002)

\item Schramm, D.N. \& Turner, M.S. 
Big-bang nucleosynthesis. Rev.\ Mod.\ Phys., 70, 303-318 (1998)

\item Fields, B.D. \& Olive, K.A. 
The Revival of Galactic Cosmic-Ray Nucleosynthesis?.
\napj, 516, 797-810 (1999)

\item Burbidge, E.M., Burbidge, G.R., Fowler, W.A., \& Hoyle, F.
Synthesis of the elements in stars.
Rev.\ Mod.\ Phys., 29, 547-650 (1957)

\item Tinsley, B.M. 
Stellar lifetimes and abundance ratios in chemical evolution.
\napj, 229, 1046-1056 (1979)

\item White, R.L., et al.\ The FIRST bright quasar survey 
II. 60 nights and 1200 spectra later. \napjs, 126, 133-207 (2000)

\item Wolfe, A.M., Turnshek, D.A., Smith, H.E., \& Cohen, R.D.
Damped Lyman-alpha absorption by disk galaxies with large redshifts 
I - The Lick survey. \napjs, 61, 249-304 (1986)

\item Sheinis, A.I., Miller, J., Bigelow, B., Bolte, M., Epps, H.,
Kibrick, R., Radovan, M., \& Sutin, B. 
SI, a New Keck Observatory Echellette Spectrograph and Imager.
\npasp, 114, 851-865 (2002)

\item Prochaska, J.X., Gawiser, E., Wolfe, A.M., Cooke, J., \&
Gelino D. The ESI/Keck~II Damped \lya Abundance Database.
\napjs, submitted

\item Vogt, S.S., et al.\ 
HIRES: the high-resolution echelle spectrometer on the Keck 10-m Telescope.
SPIE, 2198, 362-375 (1994)


\item Savage, B. D. and Sembach, K. R.
Interstellar abundances from absorption-line observations with the Hubble Space
Telescope.  \naraa, 34, 279-330 (1996)

\item Jura, M., \& York, D.G. 
Observations of interstellar chlorine and phosphorus.
\napj, 219, 861-869 (1978)

\item Woosley, S.E. \& Weaver, T.A.
The evolution and explosion of massive stars II. Explosive hydrodynamics and
nucleosynthesis. \napjs, 101, 181-235 (1995)

\item Trager, S.C., Faber, S.M., Worthey, G., \& Gonz${\rm \acute a}$lez, J.J.
The stellar population histories of local early-type galaxies
I. Population parameters. \naj, 119, 1645-1676 (2000)

\item Pettini, M., Shapely, A.E., 
Steidel, C.C., Cuby, J.-G., Dickinson, M., Moorwood, A.F.M.,
Adelberger, K.L. \& Giavalisco, M. 
The rest-frame optical spectra of Lyman break galaxies: 
Star formation, extinction, abundances, and kinematics.
\napj, 554, 981-1000 (2001)

\item Pettini, M., Ellison, S.L., Steidel, C.C., 
Adelberger, K.L., Dickinson, M., \& Giavalisco, M.
The Ultraviolet Spectrum of MS 1512-CB58: An Insight into Lyman-Break 
Galaxies. \napj, 528, 96-107 (2000)

\item Saglia, R.P., Maraston, C., Thomas, D., \& Bender, R.
The Puzzlingly Small Ca II Triplet Absorption in Elliptical Galaxies.  
\napjl, 579, 13-16 (2002)

\item Woosley, S.E., Hartmann, D., Hoffman, R.D., \& Haxton W. 
The $\nu$-process. \napj, 356, 272-301 (1990)

\item Cass\'e, M., Lehoucq, R., \& Vangioni-Flam, E. 
Production and evolution of light elements in active star-forming regions.
Nature, 373, 318-321 (1995)

\item Busso, M.; Gallino, R.; Wasserburg, G. J.
Nucleosynthesis in Asymptotic Giant Branch stars: 
relevance for Galactic enrichment and Solar System formation.
\naraa, 37, 239-309 (1999)


\item Hoffman, R.D., Woosley, S.E., Fuller, G.M., \& Meyer, B.S.
Production of the light p-process nuclei in neutrino-driven winds. 
\napj, 460, 478-488 (1996)

\item Wolfe, A.M., Prochaska, J.X., \& Gawiser, E. 
CII$^*$ absorption in damped \lya systems: A new window on
the star formation history of the universe.
\napj, submitted (2003)


\end{enumerate}

}

\end{document}